\documentclass[prl,twocolumn,showpacs,preprintnumbers,amsmath,amssymb]{revtex4}
\newcommand{\ket}[1]{\left | #1 \right\rangle}

\usepackage[dvips]{graphics}

\begin{document}

\title{When can gravity path-entangle two spatially superposed masses?}

\author{Chiara Marletto and Vlatko Vedral}
\affiliation{Clarendon Laboratory, University of Oxford, Parks Road, Oxford OX1 3PU, United Kingdom and\\Centre for Quantum Technologies, National University of Singapore, 3 Science Drive 2, Singapore 117543 and\\
Department of Physics, National University of Singapore, 2 Science Drive 3, Singapore 117542}

\date{\today}

\begin{abstract}
An experimental test of quantum effects in gravity has recently been proposed, where the ability of the gravitational field to entangle two masses is used as a witness of its quantum nature. The key idea is that if gravity can generate entanglement between two masses then it must have at least some quantum features (i.e., two non-commuting observables). Here we discuss what existing models for coupled matter and gravity predict for this experiment. Collapse-type models, and also quantum field theory in curved spacetime, as well as various induced gravities, do not predict entanglement generation; they would therefore be ruled out by observing entanglement in the experiment. Instead, local linearised quantum gravity models predict that the masses can become entangled. We analyse the mechanism by which entanglement is established in such models, modelling a gravity-assisted two-qubit gate. \end{abstract}


\maketitle                           


Witnessing quantum effects in the gravitational field has traditionally been considered extremely hard, due to the weakness of the gravitational interaction. For instance, there have been claims that detecting a graviton is practically impossible,  \cite{Dyson, Boughn}. The predictions from linearised models of quantum gravity, therefore, would be unobservable, thus making the quantisation of the gravitational field itself questionable. 

Recently, a different class of tests has been introduced, based on an indirect method \cite{MAVE, SOUG}, which is not affected by such difficulties. Instead of measuring directly quanta of the gravitational field, the proposal is to test the ability of the gravitational field to entangle two masses each in a superposition of two different locations. The logic is that if gravity can generate entanglement between the two masses, then it must be quantum. By a system "being quantum" hereinafter we shall mean that in order to describe it one needs at least two variables that do not commute with one another. This is an indirect test, because the quantum features of the field are detected by measuring observables of two test masses only (rather than, for instance, detecting gravitons or fluctuations of space-time). 

Specifically, in the experiment each mass is put into a superposition of two paths via an interferometer. If the interferometers are close enough, the state of each mass can be significantly modified by interacting gravitationally with the other mass. According to how matter couples to gravity (see below for a discussion), different outcomes may occur. However if at the output of the interferometers the two masses are entangled, one can conclude, via the argument presented in \cite{MAVE}, that the gravitational field that mediated the entanglement must itself be quantum. Thus, the entanglement between the two masses mediated by gravity is proposed as a witness of the quantisation of the field. That argument does not assume any specific dynamics \cite{MAVE}; but it assumes that there is no instantaneous action at a distance between A and B (which is why C must be the mediator). As noted in \cite{REHA}, observing entanglement would confirm some type of non-classicality in the field, in the form of there being at least two non-commuting variables, but it would not guarantee that the field is a fully-fledged quantum system. For example, one of the two non-commuting variables might be not physically observable.

This experimental proposal opens several deep questions, some of which we shall address in this manuscript. There are, at present, a number of proposed models for coupling matter with gravity. In this paper, we analyse which predictions they provide for the experiment in question. We restrict attention to three cases: the linearised and canonical approaches \cite{Kiefer}; collapse-type models, \cite{Penrose, Diosi, GRW} and semiclassical gravity, \cite{DEWITT}. 

From our analysis it emerges that both the collapse models, and also semiclassical approaches to gravity, would be ruled out if entanglement were to be observed, because in such models the gravitational field is classical - it only has one observable, thus it would not be capable of generating any entanglement between the masses. Linear quantum gravity (to which all the canonical approaches would have to reduce) would, on the other hand, be confirmed. Thus this experiment differs from previous experiments that couple a quantum system to gravity, such the COW and related experiments \cite{COW,Brukner}, where the gravitational field remains completely classical. 

Using linearised quantum gravity, we also explain how the entanglement between the two masses is generated, by using the gravitational field as a third system that acts as a mediator between the two. We shall discuss specifically the amount of entanglement between the masses and the field, as well as the entanglement between the two masses. The observed effects in this experiment are not general-relativistic; confining attention to the Newtonian contributions suffices to explain the proposed experiment. This therefore suggests that gravity ought to be quantised even if it was only Newtonian (provided that there is no instantaneous action at a distance). In this regime the non-perturbative approaches to quantum gravity, such as loop quantum gravity \cite{ROV} and string theory \cite{STR}, agree with the predictions of the linearised approach; thus the proposed experiment would test their prediction, too, in this regime. 

We first recall the logic of the experiment. Consider two equal masses $Q_1$ and $Q_2$ each in, say, two Mach-Zehnder interferometers, each located horizontally in the Earth's gravitational field (so that both masses are approximately subject to the same background field, which can therefore be neglected for the purpose of computing phase differences). 
Let one of the arms be indicated by $0$ and the other by $1$.  Each mass is after the first beam-splitter is in the state $\frac{1}{\sqrt 2} \left(\ket{0}+\ket{1}\right )$. The distance from one arm of the interferometer to the closer arm of the second interferometer is $r_1$; the distance to the other arm of the second interferometer is $r_2$. If one supposes that the masses on different paths interact via the gravitational field, different things can happen according to how matter and gravity couple.  Broadly speaking, the superpositions might undergo some collapse or not; and if they do not, they may or may not become entangled.

For clarity, we first explain how to realise a two-qubit entangling gate acting on the masses, by way of letting them interact separately with the gravitational field, but not directly with one another. 
The two masses only get entangled via gravity, and do not interact directly (thereby ruling out action at a distance). We shall model this by requiring that the Hamiltonian does not contain products of operators acting both on $Q_1$ and $Q2$.

Consider a simple model where the gravitational field $C$ is treated as a single quantum harmonic oscillator (or, more accurately, a collection of them). In the linearised model of quantum gravity, $a$ and $a^{\dagger}$ can be interpreted as the annihilation and creation operators for gravitons. The two masses $Q_1$ and $Q_2$ can be initially modelled as two qubits -- whose z-component represents a discretised position of each mass. (In our earlier discussion, the position represents one of two arms of a Mach-Zehnder interferometer, but it could be more general.) Its eigenstates $\ket{a}$ where $a\in \{0,1\}$ represent the situation where the mass is in a definite position $a$; the state $\ket{ab}$ describes the situation where the first mass is in position $a$ and the second in position $b$. 

We shall now analyse how the relative phases in the quantum superpositions of masses are established during the double interference experiment, thus inducing entanglement. First the masses get entangled to the field, then the phases are generated through a generalised controlled-phase gate.

Immediately after the action of the first beam splitter, the state of the two masses and the field is
$\ket{\phi_0}=\frac{1}{2}\sum_{a,b\in\{0,1\}}\ket{ab}\ket{\alpha}$,
where $a, a^{\dagger}$ are the bosonic creation and annihilation operators for the field; and $\ket{\alpha}={\rm e}^{-\frac{1}{2}|\alpha|^2}\exp(\alpha (a^{\dagger}-a))\ket{0}$ is a coherent state representing the spatial modes of gravity - possibly a continuum. 

The two masses and the field then evolve into the state 
\begin{eqnarray}
\ket{\phi_{E1}}=U_1\ket{\phi_0}=\frac{1}{2}\sum_{a,b\in\{0,1\}}\ket{ab}\ket{\alpha_{a,b}}
\end{eqnarray}
where $U_1\doteq \sum_{a,b\in\{0,1\}}P_{ab}\otimes D(\xi_{ab})$ and $\ket{\alpha_{a,b}}=\ket{\alpha+i\sqrt{\xi_{ab}}}=D(\xi_{a,b})\ket{\alpha}\;$; we have defined the displacement operator as $D(\xi_{a,b})=\exp{(i\sqrt{\xi_{a,b}}(a^{\dagger}-a))}$ with $\sqrt{\xi_{ab}}$ being a real-numbered shift that depends on the coupling between the field and the masses, that brings about the desired phase-shift $\phi_{a,b}$ at the end (see below for more details). We have also defined the projectors $P_{ab}=P_a\otimes P_b$, with $P_{0,1}=\frac{(id\pm\sigma_z)}{2}$ being the projector operator for the location of each mass; and $w$ is some real number with the property that $w\xi_{a,b}=\phi_{a,b}$. We have also assumed that establishing the entanglement
between the field and the masses takes place on time-scales much faster than the process that transfers the phase $\phi_{a,b}$ back from the field to the masses, evolving their composite system to the state $\ket{\phi_{E2}}=U_2\ket{\phi_{E1}}\approx \frac{1}{2}\sum_{a,b\in\{0,1\}}\exp{(i\phi_{a,b})}\ket{ab}\ket{\alpha_{a,b}}\;$, where $U_2=\exp(w(a^{\dagger}a))$ and we, for the sake of this simple illustration, we have assumed to be in the regime where $|\alpha|$ is large and real (later the full linearised model will present the exact solution for any coherent state).  Finally, the interaction $U_1^{\dagger}$ between the field and the masses brings the field back to its original state and the masses remain entangled (to the degree depending on the phase): $\frac{1}{2}\sum_{a,b\in\{0,1\}}\exp{(i\phi_{a,b})}\ket{ab}\ket{\alpha}\;.$

The key fact is that the above process relies on two complementary observables of the field. This can be seen by noticing that the observables $\frac{1}{2i}(a-a^{\dagger})$ and $a^{\dagger}a$ are needed to generate the unitaries $U_1$ and $U_2$. The values $w$ and $\xi_{ab}$ depend on the details of the interaction between gravity and the masses.

The entanglement between the field and masses can be quantified by the reduced entropy of the masses, and it is independent of $w$. Since the field and the masses are weakly entangled, a good approximation of the reduced entropy is the linear entropy ($S_L=1-{\rm Tr(\rho_{Q_1,Q_2}^2)}$, where $\rho_{Q_1,Q_2}$ is the reduced state of the two masses). The magnitude of the reduced entropy is given by one minus the overlap between the two gravitational states squared as in: $1- |\langle\alpha_{ab}|\alpha\rangle|^2 = 1- \exp{(-\xi_{ab})} \approx \xi_{ab}\; .$ This quantity could be very small compared to one, while still generate the desired entanglement between the two masses. Assuming Newtonian gravity, one has: $\phi_{ab}=w\alpha_{ab}=\frac{Gm^2}{\hbar d_{ab}}\Delta t=\left (\frac{m}{m_P}\right )^2\frac{c}{d_{ab}}\Delta t\;$,
where $\Delta t$ is the interaction time between the two masses, $m_{P}$ is Planck's mass and $d_{ab}$ is the distance between the position $a$ of the first mass and position $b$ of the second mass. One can identify $\alpha_{ab}=\left (\frac{m}{m_P}\right )^2$. The entanglement between a spatially superposed mass and the gravitational field (if, indeed, it is quantum) would then offer another way of understanding the Planck mass. Namely, if we really want a spatially superposed mass to entangle appreciably to the surrounding gravitational field, according to the above formula, we need to engage masses on the order of and larger than the Planck mass. The full linearised model, to be presented below, leads to the same conclusion.

Supposing that only the closer arms of the interferometers, labelled as 1, interact, the only phase present in the state of the masses before they enter their respective final beamsplitter would be $\phi_{11}$. In each of the interferometers, the probability for the mass to emerge on path $0$ is $p_0=\frac{1}{2}\left(1+\cos^2{\frac{\phi_{11}}{2}}\right)$ (and $p_1=1-p_0$). When the two masses are maximally entangled by the action of the gravitational field, in which case $p_0=p_1=\frac{1}{2}$. This happens when $\phi_{11}={\pi}$. On the other hand, when the two masses are not entangled they undergo, separately, an ordinary interference experiment: that happens when $\phi_{11}=2n\pi$. In that case, each mass emerges on path $0$ of the interferometer. For a fixed mass, by varying the arms' distance or their length, it is in principle possible to interpolate between those two cases, thus demonstrating entanglement. To confirm entanglement, one would require to measure two complementary observables on each of the masses. For example, let us focus on the maximally entangled state. It is of the form $|0\rangle |+\rangle +
|1\rangle |-\rangle$. Therefore if we measure the path of the first particle (the effective Pauli Z measurement if we think of the path as a qubit), the second mass interferes with either the plus or the minus phase (which means that it is in an eigenstate of the Pauli X). If on the other hand, the second mass is first measured in X, there is no interference of the first mass (meaning that $X_1$ and $X_2$) are not correlated. Therefore the observable $X_1Z_2 + Z_1X_2$ will suffice to witness entanglement. 

We now explain how to perform the calculation of the phases in the proposed experiment using linearised quantum gravity. For convenience, we adopt the Hamiltonian picture (as in the analogue problem in quantum optics). The evolution generated by this Hamiltonian can be approximated in discrete steps by the gate model presented above. 

The linearised gravity-matter interaction Hamiltonian contains the complementary observable $\frac{1}{\sqrt{2}}(a+a^{\dagger})$. This is obtained from the general linearised Hamiltonian, \cite{Boughn}:
\begin{equation}
H^G_{int} =- \frac{1}{2} h_{\mu\nu} T^{\mu\nu}
\end{equation}
where $T^{\mu\nu}$ is the stress-energy tensor and $h_{\mu\nu}$ is the perturbation of the metric tensor $g_{\mu\nu}$ away from the flat (Minkowski) spacetime. The quantised gravitational field is then written in terms of the graviton creation and annihilation operators as:
\begin{equation}
h_{\mu\nu} \propto \sum_{\sigma} \int \frac{d^3k}{\sqrt{\omega_k}} \{ a(k,\sigma)\epsilon_{\mu\nu} (k,\sigma) e^{ik_\lambda x^\lambda} + h.c.\}
\end{equation}
where $\epsilon_{\mu\nu}$ is the polarisation tensor, $\sigma$ indicates two non-vanishing gravitational polarisations, while $\omega_k$ and $k$ represent the frequency and wavenumber of the mode respectively (we are using the Einstein's convention of summation). 

In our experiment, the masses are non-relativistic and the stress-energy tensor would simplify to $T_{00} = m$. We can also consider, for simplicity, a single polarisation and a discrete sum over the relevant gravitational quantum modes.
The total Hamiltonian involving two masses and the gravitational field is therefore
\begin{eqnarray}
H & = & mc^2 (b^{\dagger}_1 b_1 +b^{\dagger}_2 b_2) + \sum_k \hbar \omega_k a^\dagger_{k}a_{k} \\
& - &   \sum_{k,  n\in \{1,2\}} \hbar g_k  b^{\dagger}_n b_n (a_{k}e^{i k x_n}+a^\dagger_{k}e^{-i k x_n})
\end{eqnarray}
where the first two terms are the free Hamiltonians of the masses and the field respectively. 
We assume that the gravitation-matter coupling constant is given by
$g_k = mc \sqrt{\frac{2\pi G}{\hbar \omega_k V}}\;$, where $V$ is the relevant volume of quantisation (which will not feature in the relevant observables). The evolution of two stationary masses of value $m$ at positions $x_1$ and $x_2$ and the initial gravitational vacuum state can be solved exactly: $
e^{iH t} |m\rangle |m\rangle |0\rangle =\exp\{\hbar \sum_kV(k)t\} |m\rangle |m\rangle |\sum_k \frac{g}{\omega_k}(e^{-ikx_1}+e^{ikx_2})\rangle$, 
where $V(k)= \frac{g_k^2}{2\omega_k}(1+2\cos(-ik(x_2-x_1)))$.

When acting on the initially superposed state, this generates entanglement between the two masses by implementing the controlled-phase gate $U_1U_2$ described in the previous section. Note that only the position-dependent part of $V(k)$ contribute to the phase difference.
The continuum version of the position-dependent part of $V(k)$, obtained by replacing the sum over $k$ by an integral, is:
\begin{equation}
{\rm Re} \left \{V \int dk \frac{4\pi Gm^2}{\hbar k^2 V}e^{-ik(x_1-x_2)}\right \} = \frac{Gm^2}{\hbar (x_2-x_1)}
\end{equation}
which gives us the same phase as the particles coupled via the Newtonian potential, as described in the gate model.  

We can assume that the interaction between the masses and the field is 'elastic', i.e., when the two masses are brought back to their original state, where each one of their positions is sharp, the field goes back to the original state, and it is unentangled with the masses. However, even if the interaction were not perfectly elastic, since the entanglement between the field and the masses is very small (for masses below Planck mass), as computed earlier, the state of the two masses is approximately not entangled with the field at the end, thus leaving the field approximately unchanged. 
The same result can be obtained with the usual Lagrangian formulation of quantum field theory, where the interaction is established via the exchange of a single graviton between the two masses and the field.

A class of theories that would be refuted by observing entanglement in the proposed experiment are the so-called 'semiclassical' theories of gravity, \cite{DEWITT}, \cite{DAV}. In these theories, the background spacetime is classical, but the back-action of the masses prepared in some quantum state on the field can be taken into account as an average of the energy-momentum tensor in the quantum state of the masses.  The Einstein's equation reads: $R_{\mu, \nu}-\frac{1}{2}Rg_{\mu, \nu}=8\pi G\langle T^{\mu\nu}\rangle$, 
where $R_{\mu,\nu}$ is the Ricci tensor; $R$ is its trace; and $g_{\mu, \nu}$ is the metric tensor.

These theories have provided powerful predictions such as the Unruh effect and the Hawking radiation, \cite{DAV}. Yet, they cannot adequately describe quantum effects in gravity, as already pointed out in \cite{PAGE}. This is because they resort to a field which is classical - in the sense that it has no pair of non-commuting observables. The field strength at each point is determined by an average of the stress-energy tensor in the quantum state of the masses. 

In the situation of our experiment, each mass would therefore be affected by the average of the gravitational field generated by the other superposed mass. Supposing each mass is initially in an equally-weighted superposition of the two possible locations, each mass would experience the potential generated by the other mass $m$ positioned at a distance which is the average of the position of the other mass in its quantum state. 

Assuming once more that only the gravitational interaction affects the branch corresponding to the arm of the interferometer closer to the other (labelled by $1$), the state of the mass would become $\frac{1}{\sqrt{2}}(|0\rangle+ e^{i\phi_m}|1\rangle )$, where $\phi_m=G\frac{m^2t}{\hbar d_m}$ where $d_m=\frac{d_1+d_2}{2}$. Likewise, by symmetry, for the other mass. Thus the phase acquired would be a local phase, which cannot generate entanglement between the masses. Each mass would be undergoing a separate, COW-type experiment: the state of the two masses would be at any time a product state. Semiclassical theories would therefore be refuted by observing entanglement in the proposed experiment. The same prediction of no entanglement would be reached by models that resort the (non-equivalent) procedure of averaging the linearised quantum gravity Hamiltonian in the quantum state of the two masses. This would also provide only local phases (albeit different from the former case).

The other class of theories that would be ruled out by witnessing entanglement in this experiment are all those collapse-type models which would predict a collapse of the wave-function of each mass at the scales of the experiment \cite{Diosi,Penrose,GRW}. For masses such as those proposed in our experiment, such models would predict that each of the masses undergoes an irreversible transition to a state where the position is sharp, which means that no entanglement could be generated via the gravitational interaction. Consider for example the decoherence time of a superposition of a mass of $10^{-12}$ kg across two different locations, at a distance of approximately $10^{-4}$m, which could be the spatial extent of each interferometer. The decoherence time according to Penrose's collapse models, \cite{Penrose} would be of the order $t=\frac{\hbar}{G\frac{m^2}{d}}\approx 10^{-13}s$, well below $10^{-6}$s, required for our experiment.  Note that there is a subtle difference between collapse occurring, and decoherence happening while an interference experiment is taking place. For example, when a neutron undergoes interference, its spin couples to neighbouring spins and affects their state. Since in the interferometer the neutron exists in a superposition of 
two different spatial locations, the neighbouring spins are in two different states which are entangled with the spatial states of the neutron. However, when the two arms of the interferometer are recombined to measure the interference the two
environmental states effectively become the same. This is why despite the fact that the neutron was fully entangled with environmental spins inside the interferometer we are still able to observe interference at the end. A massive superposition can also become entangled with the gravitational field, as we explained earlier, and still undergo a coherent evolution. It is this kind of effect that could be discriminated from the genuine collapse that would occur according to collapse models. We can aslo discriminate de-phasing induced by gravity from the gravitationally induced collapse, because the former would still allow generation of entanglement, while the latter would not. However, we cannot discriminate gravitationally induced spontaneous emission (still a fully quantum effect due to the vacuum state of the quantised gravitational field) from a collapse, because in both cases the interference pattern would be destroyed. 

Finally, another approach to gravity is that proposed in \cite{SAK, REV}, which treats classical gravity as an induced field by the quantum vacuum fluctuations of all other fields. According to this logic, gravity is not a fundamental force and therefore need not be quantised at all. Given that the gravitational field obtained via these approaches is semiclassical, they are not, in the present form, able to account for entanglement generation in our experiment. Still, it may be possible that entanglement generation is explained as caused by other quantum fields, leading to an effective gravitational phase. This remains an open question.

\textit{Acknowledgments}: CM thanks the Templeton World Charity Foundation and the Eutopia Foundation. VV thanks the Oxford Martin School, the John Templeton Foundation, the EPSRC (UK). This research is also supported by the National Research Foundation, Prime Minister's Office, Singapore, under its Competitive Research Programme (CRP Award No. NRF- CRP14-2014-02) and administered by Centre for Quantum Technologies, National University of Singapore.

\end{document}